\RequirePackage{fix-cm}
\documentclass[smallextended]{svjour3}       % onecolumn (second format)
\smartqed  % flush right qed marks, e.g. at end of proof
\usepackage{graphicx}

\usepackage{mathptmx}      % use Times fonts if available on your TeX system

\usepackage{siunitx}

\begin{document}

\title{Two-Photon Exchange: Future experimental prospects
}


\author{Jan C. Bernauer
}


\institute{J.~C. Bernauer \at
              Massachusetts Institute of Technology\\
              \email{bernauer@mit.edu}
}

\date{Received: date / Accepted: date}

\maketitle

\begin{abstract}
The proton elastic form factor ratio is accessible in unpolarized Rosenbluth-type experiments as well as experiments which make use of polarization degrees of freedom. The extracted values show a distinct discrepancy, growing with $Q^2$. Three recent experiments tested the proposed explanation, two-photon exchange, by measuring the positron-proton to electron-proton cross section ratio. In the results, a small two-photon exchange effect is visible, significantly different from theoretical calculation. Theory at larger momentum transfer remains untested. This paper discusses the possibilities for future measurements at larger momentum transfer.
\end{abstract}

% Head 1
\section{Introduction}
\begin{figure}[htb]
  \centerline{\includegraphics[width=\textwidth]{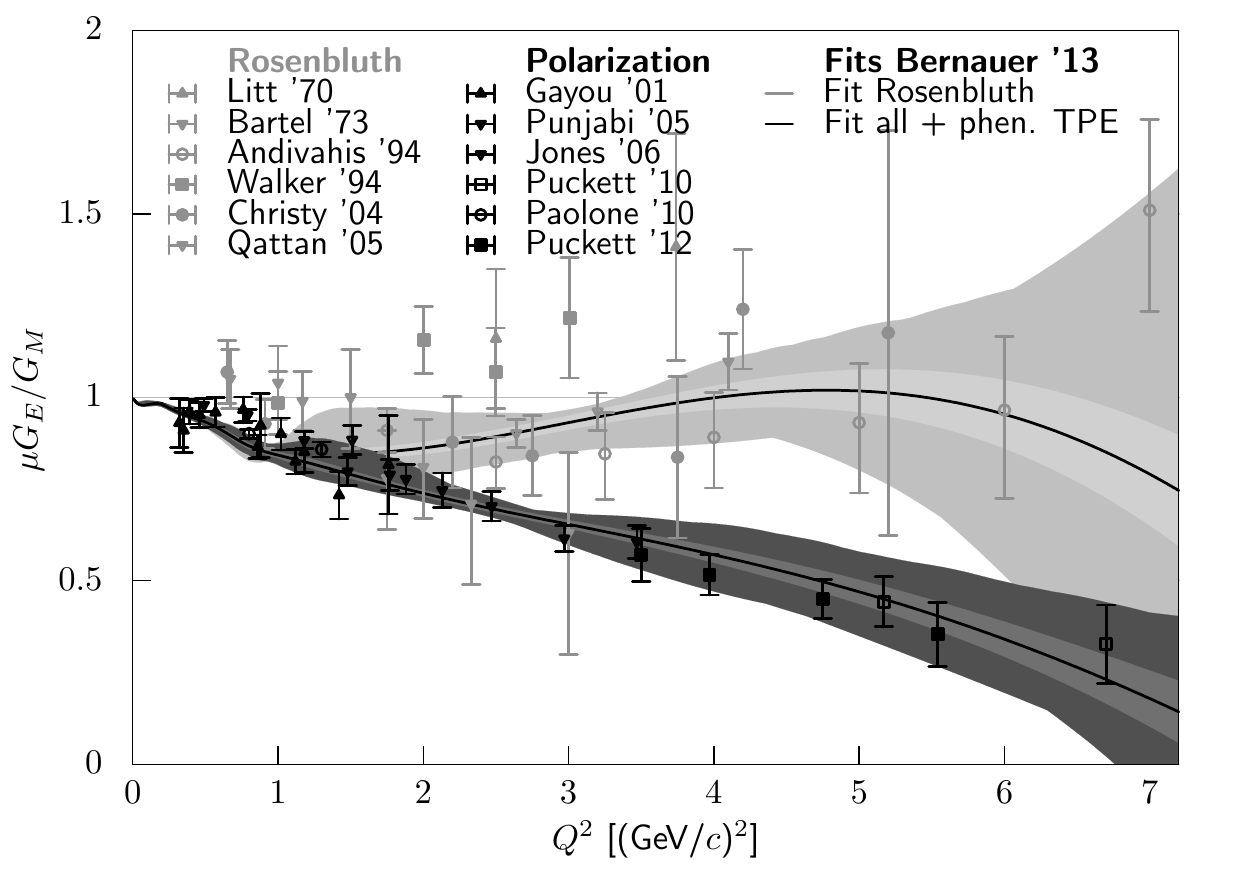}}
  \caption{\label{figratio}The proton form factor ratio $\mu G_E/G_M$,  determined via Rosenbluth-type (gray points, from \cite{litt,bartel,andivahis,walker,christy,qattan}) and polarization-type (black points, from \cite{gayou,punjabi,jones,puckett10,paolone,puckett12}) experiments. While the former show a constant ratio, the latter indicate a linear downward trend. Curves represent phenomenological fits \cite{bernauer13}, to either the Rosenbluth-type world data set alone (light gray curves) or to all data (dark gray curves).}
\end{figure}

Proton elastic form factors have been studied intensively with electron-proton scattering using unpolarized beams and target. The experiments produced data over an extensive range of (negative) four-momentum transfers, $Q^2$. Via the so-called Rosenbluth separation technique, the two elastic form factors were separated. More recently, experiments exploiting the polarization of beam or target measured the form factor ratio directly. While the former see rough agreement with scaling, i.e., a more or less constant ratio even for large $Q^2$, the latter show a roughly linear fall-off of the ratio. Figure \ \ref{figratio} shows a selection of the available data and recent fits.

The form factors encode the distribution of charge and magnetization in the proton and this "form factor ratio puzzle" is a limiting factor in their precise determination. It is therefore of importance to resolve this puzzle.

\section{Two-photon exchange}
Blunden et al.\ \cite{Blunden:2003sp} suggested that hard two-photon exchange (TPE), neglected in standard radiative corrections, could be an important effect in Rosenbluth-type experiments, and that an inclusion of TPE might resolve the discrepancy. Two-photon exchange corresponds to a group of diagrams in the second order Born approximation of lepton scattering, namely those where two photon lines connect the lepton and proton. While the ``soft'' case, when one of the photons has negligible momentum, is included in the standard radiative corrections, like ref.\ \cite{MoTsai,Maximon2000}, to cancel infrared divergences from other diagrams, the ``hard'' part, where both photons can carry considerable momentum, is not. The exact division in ``soft''and ``hard'' is arbitrary and depends on the specific radiative correction used.

\subsection{Theoretical calculations}
Current theoretical calculations can be roughly divided into two groups: hadronic calculations, e.g.\ \cite{Blunden:2017nby}, which are believed valid for $Q^2$ from 0 up to a couple of GeV$^2$, and GPDs based calculations, e.g.\ \cite{Afanasev:2005mp}, valid from a couple of GeV$^2$ and up. 

\subsection{Phenomenological extraction}
The amount of data available for the form factor ratio allows for an extraction of the expected TPE size. In \cite{bernauer13}, the authors built a model based on the following assumptions:
\begin{itemize}
\item TPE is the dominany source of the difference.
\item TPE affects the Rosenbluth-type experiments and leaves polarization data unchanged. This is good approximation as the effect of TPE on the cross section is magnified in the Rosenbluth separation to a substantially larger effect on $G_E$ for $Q^2>>0$.
\item The effect is roughly linear in $\epsilon$. This is supported by the fact that no strong deviations from a straight line have been found in Rosenbluth separations so far.
\item The effect vanishes for forward scattering, i.e., for $\epsilon=1$.
\item For $Q^2\rightarrow 0$, TPE is given by the Feshbach Coulomb correction \cite{McKinley:1948zz}. Modern theoretical calculations have the same limit.
\end{itemize}
Assuming a correction of the form $1+\delta_{TPE}$ to the cross section, with 
\begin{equation}
\delta_{TPE}=\delta_\mathsf{Feshbach}+a(1-\epsilon)\ln{(1+b*Q^2)},\label{eqfesh}
\end{equation}
the authors could fit the combined world data set with excellent $\chi^2$. This extraction will be used in the following to predict the size of the effect.

\section{Current status}
Three contemporary experiments have tried to measure the size of TPE, based at VEPP-3 \cite{Rachek:2014fam}, Jefferson Lab (CLAS, \cite{Adikaram:2014ykv}) and DESY (OLYMPUS, \cite{Henderson:2016dea}). The next-order correction to the elastic lepton-proton cross section contains terms corresponding to the product of the diagrams of one-photon and two-photon exchange. These terms change sign when switching between $e^-$ and $e^+$. Therefore, the size of TPE can be determined by measuring the ratio of positron to electron scattering: $R_{2\gamma}=\frac{\sigma_{e^+}}{\sigma_{e^-}}\approx 1+2\delta_{TPE}$.

\begin{figure}[htb]
  \centerline{\includegraphics[width=0.5\textwidth]{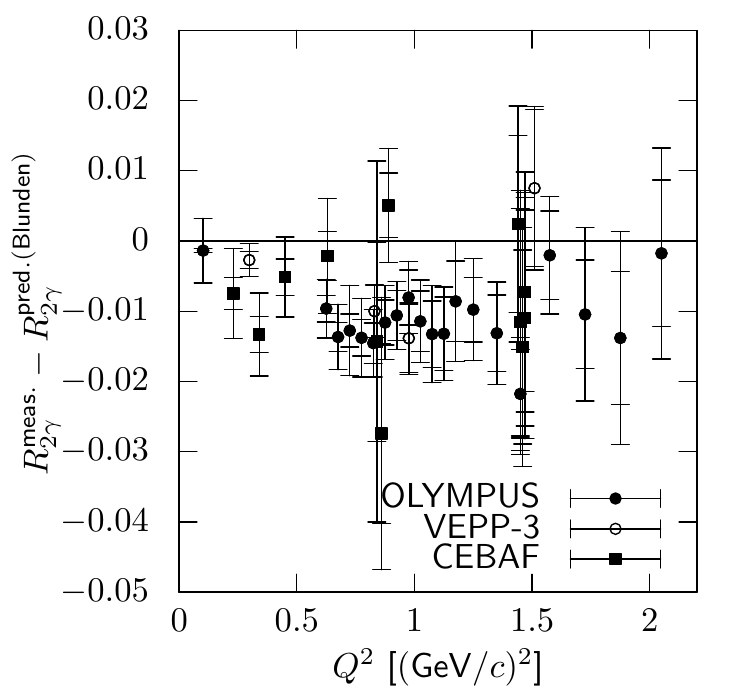}\includegraphics[width=0.5\textwidth]{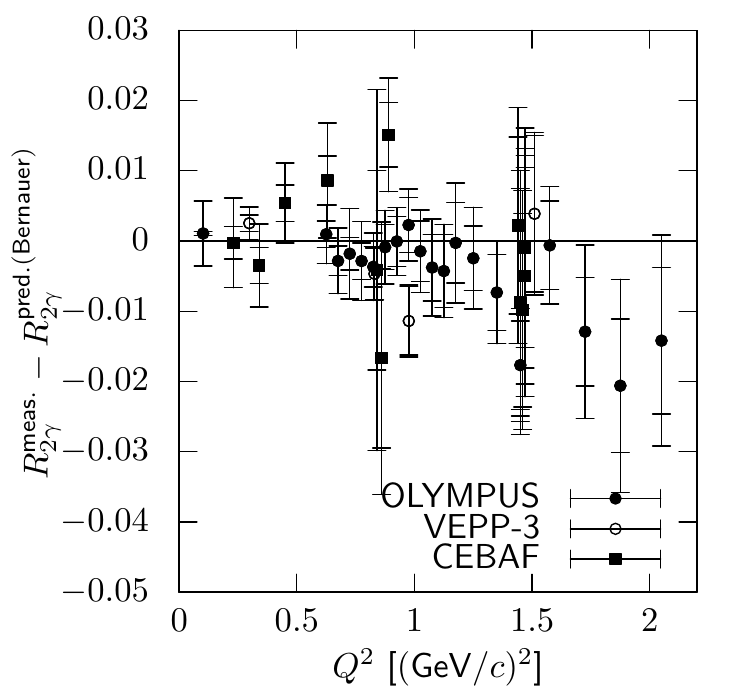}
  }
  \caption{\label{figdiff} Difference of the data of the three recent TPE experiments \cite{Rachek:2014fam,Adikaram:2014ykv,Henderson:2016dea} to the calculation in \cite{Blunden:2017nby} (left) and the phenomenological prediction from \cite{Bernauer:2013tpr} (right).}
\end{figure}

In Fig.\ \ref{figdiff} the difference of the data of the three experiments to the calculation by Blunden et al.\ \cite{Blunden:2017nby} and the phenomenological prediction by Bernauer et al.\ \cite{Bernauer:2013tpr} is shown. The three data sets are in good agreement which each other, and appear about 1\% lower than the calculation. The prediction appears closer for most of the $Q^2$ range, but is above the data for the largest available $Q^2$. This is worrisome, as this coincides with the opening of the divergence in the fits in Fig.\ \ref{figratio}. It might be an indication for an additional effect beyond TPE driving the discrepancy.

No hard TPE is ruled out by the data. The experiments agree with the phenomenological prediction with a reduced $\chi^2$ of 0.68. Compared to that, the theoretical calculation (red.\ $\chi^2$ of 1.09) is significantly worse, and the large normalization shifts to achieve this value rules them out at 99.6\% confidence level. 

The existing data show that TPE exists, but is small the in the covered region. Hadronic calculations are close, but can not explain the data perfectly. The calculations based on GPDs are only valid at higher $Q^2$ and are so far not tested by any experiment.

For a more in-depth review, see \cite{Afanasev:2017gsk}. Without a resolution of the puzzle and a test of TPE at larger $Q^2$, the extraction of reliable form factor information is impossible,  especially from the high precision, large $Q^2$ measurements which are part of the Jefferson Lab 12 GeV program. Clearly, new data are needed. In the following, we will discuss experimental possibilities.

\section{Future experiments}
\subsection{Effect size and figure of merit}
As can be seen from Eq.\ \ref{eqfesh}, the size of TPE scales linearly with $1-\epsilon$, but only weakly with $Q^2$. The strongest signal is therefore at small $\epsilon$ and large $Q^2$. The cross section, however, drops fast exactly for the same kinematic conditions. We can construct a figure of merit to find the optimal kinematics: the ratio of expected deviation of $R_{2\gamma}$ from 1 and the expected uncertainty.
\begin{equation}
  FOM=\frac{\left|R_{2\gamma}-1\right|}{\sqrt{\Delta^2_\mathsf{stat.}+\Delta^2_\mathsf{syst.}}}
  \end{equation}
Here, the total uncertainty is split into a statistical and a systematical part. For the following, we assume a 1\% systematic error.

Positron beams of the relevant energies are rare. Two possible sites for such an experiment are DESY and Jefferson Lab.

\subsection{Measurement at Jefferson Laboratory}
Jefferson Lab is evaluating the construction of a positron source for CEBAF. We assume that such a source would enable CEBAF to deliver up to \SI{1}{\micro\ampere} of unpolarized positrons impinging on a 10 cm liquid hydrogen target, which yields an instantaneous luminosity of $\mathcal{L}=\SI{2.6}{\per\pico\barn\per\second}$.

\begin{figure}[htb]
  \centerline{\includegraphics[width=0.5\textwidth]{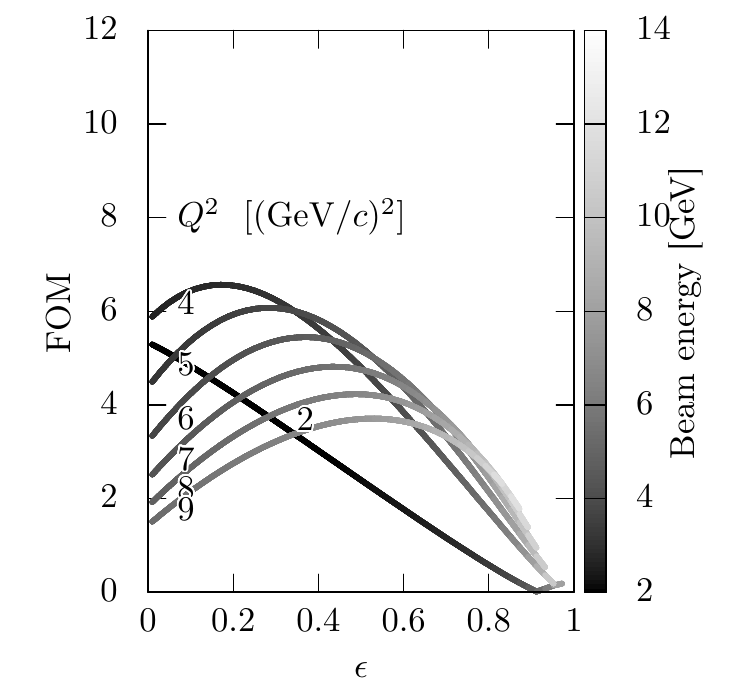}\includegraphics[width=0.5\textwidth]{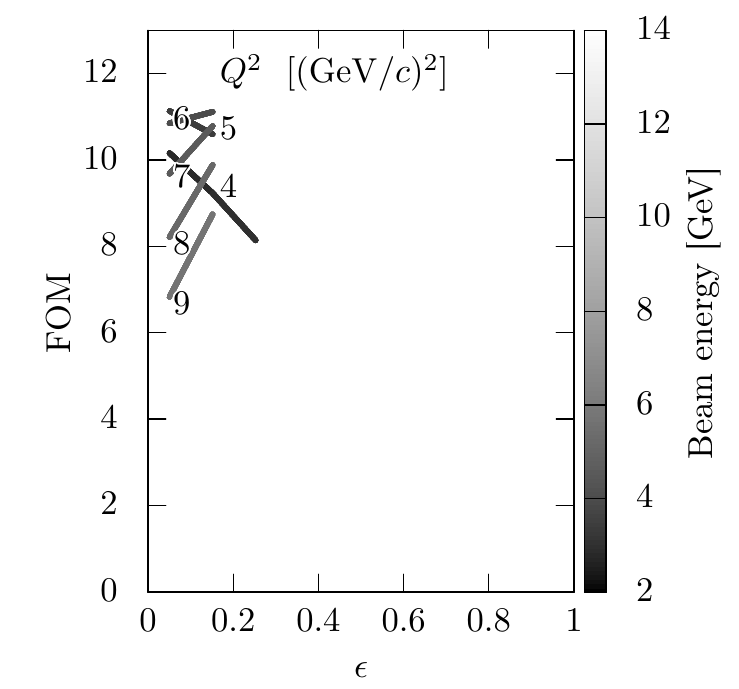}
  }
  \caption{\label{figfomjlab}Figures of merit as a function of $\epsilon$, for various $Q^2$, for days of beam per species at Jefferson lab. Left: small acceptance spectrometers, right: BigBite.}
\end{figure}

For this paper, we investigated the measurement possibilities of Hall A and C. The main spectrometers of Hall A and the HMS spectrometer in Hall C can easily be used. SHMS in Hall C is limited to forward angles, and thus $\epsilon\approx1$, if used as a lepton spectrometer, but could be used to detect the protons instead. BigBite in Hall A is limited in the maximum momentum and thus minimal angle.  However, because of  the large acceptance, measurements at very low values of $\epsilon$ are possible. Figure \ref{figfomjlab} shows the figure of merit for two days of beam per species, with the smaller-acceptance spectrometers represented by the left figure and BigBite by the right figure.

\subsection{Measurement at DESY}
\begin{figure}[htb]
  \centerline{\includegraphics[width=0.5\textwidth]{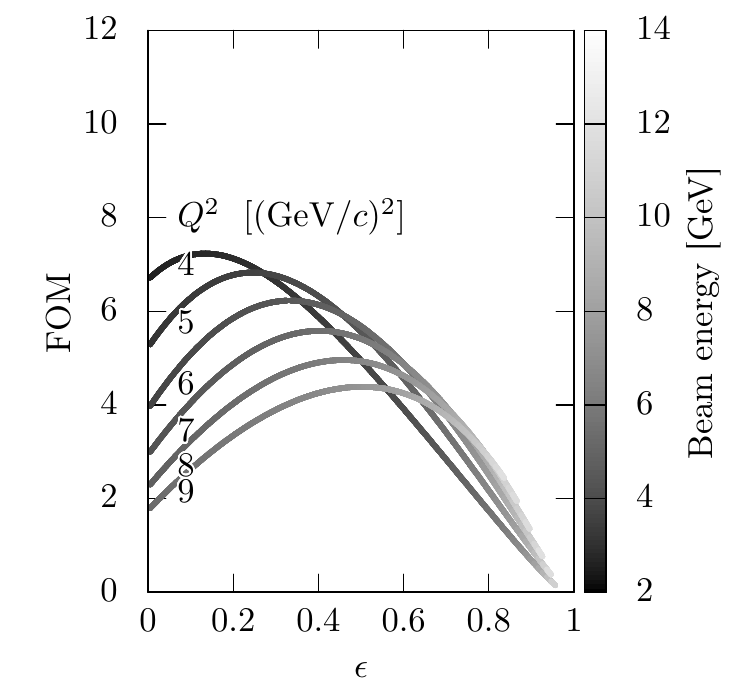}\includegraphics[width=0.5\textwidth]{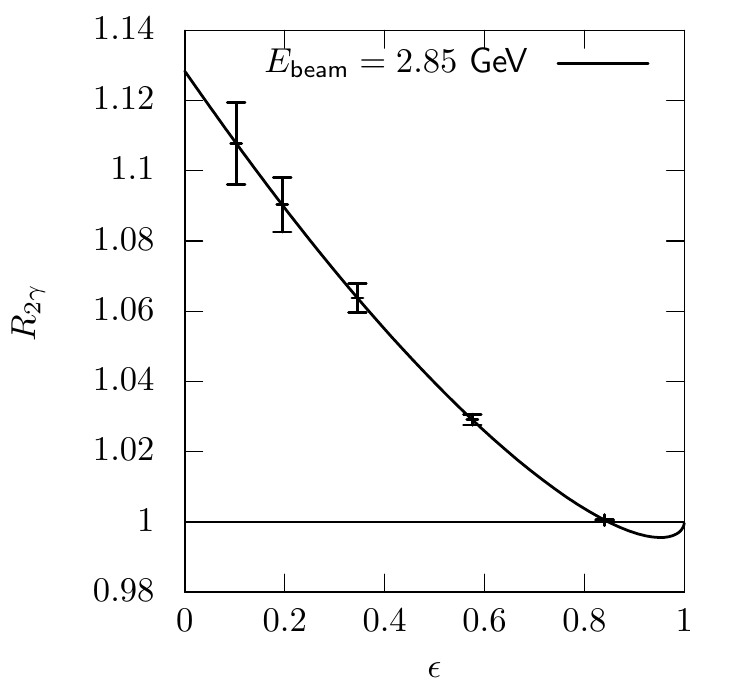}
  }
  \caption{\label{figdesy}Left: figure of merit as a function of $\epsilon$, for various $Q^2$, for 30 days of beam per species at DESY. Right: expected statistical error of data points and predicted effect size. }
\end{figure}

DESY currently investigates a new test beam facility which would make TPE measurements with a \SI{60}{\nano\ampere} beam possible. The proposed facility size and schedule constraints indicate non-magnetic calorimetric detectors as ideal, such as the those designed and built for PANDA. We assume five detector elements covering 10 msr each. The beam impinges on a 10 cm liquid hydrogen target. The left part of Fig.\ \ref{figdesy} shows the FOM plot for 30 days per species. With a 2.85 GeV beam, the experiment could test TPE up to a $Q^2$ of about 6 GeV$^2$ with more than 5$\sigma$. The projected errors for such a measurement are shown on the right.

\section{Conclusion}
The discrepancy in the form factor ratio is a serious limitation in the exact determination of the proton form factors and must be studied further in a dedicated program. The proposed test beam area at DESY could host an experiment to investigate TPE at larger momentum transfers on a short time scale. At Jefferson Lab, an upgraded CEBAF would make more precise experiments at even larger momentum transfers possible. This would test both hadronic and GPD-based theoretical calculations of TPE. Even if both calculations would be found lacking, the data would allow a phenomenological model precise enough to analyze contemporary and future form factor measurements.

% Acknowledgement
\section{Acknowledgments}
This work was supported by the Office of Nuclear Physics
of the U.S. Department of Energy, grant No.\ DE-FG02-94ER40818.

% References

%\nocite{*}
\bibliographystyle{spphys}%
\bibliography{nstar}%

\end{document}